# Geodesic photon coupling and non-Hermitian exceptional point of microcavities on topologically curved space

Yongsheng Wang, Xiaoxuan Luo, Bo Li, Zaoyu Chen, Zhenzhi Liu, Fu Liu, Yin Cai, Yanpeng Zhang, and Feng Li[*]

*Key Laboratory for Physical Electronics and Devices of the Ministry of Education & Shaanxi Key Lab of Information Photonic Technique, School of Electronic Science and Engineering, Faculty of Electronic and Information Engineering, Xi'an Jiaotong University, Xi'an 710049, China*

\* Corresponding author: [felix831204@xjtu.edu.cn](felix831204@xjtu.edu.cn)

**Abstract**

Asymmetric microcavities supporting Whispering-gallery modes (WGMs) are of great significance for on-chip optical information processing. We define asymmetric microcavities on topologically curved surfaces, where the geodesic light trajectories completely reconstruct the cavity mode features. The quality factors of the lossy chaotic and island modes in flat cavities can be increased by up to ~200 times by the space curvature. Strong and weak coupling between modes of very different origins occur when the space curvature brings them into resonance, leading to a fine tailoring of the cavity photon energy and lifetime. Finally, we prove that by varying the overall loss, an exceptional point can be clearly observed at which both the cavity photon energy and lifetime become degenerate. Our work is at the crosspoint of optical chaotic dynamics, non-Hermitian physics and geodesic optical devices, and would open the novel area of geodesic microcavity photonics.

**Introduction**

Optical microcavities supporting whispering gallery mode (WGM) are resonators confining light via successive internal reflections that are extensively investigated for applications in on-chip microlasers[1,2], sensors[3,4], filters[5], isolators[6], frequency combs[7] and gain-loss featured optoelectronic devices[8,9]. The cavity periphery basically exhibits a circular or polygonal shape in forms of microdisks, microspheres, microtoroids, microwires and microtubes[10–16]. To achieve particular features such as unidirectional lasing, asymmetric cavities were designed and fabricated by either abruptly or smoothly deforming the cavity periphery[17,18], which leads to stable and chaotic optical modes[19–21]. The chaotic motion of photons in the cavity are usually analyzed by ray dynamics which records the reflecting angle as a function of the reflecting position in a diagram called the Poincare surface of section (PSOS)[22–25], and then compared with its counterpart of simulations using wave optics. It is shown that the chaotic modes in deformed cavities are particularly useful to achieve broadband resonance and nonlinear effects such as frequency combs[22,26,27] Meanwhile, controlling the gain and loss of the microcavity modes leads to non-Hermitian physics, which provides a new mechanism for improving device performances.[4,5,21,28–30].

Recently, curved membranes are developed to form three-dimensional (3D) on-chip photonic structures, which would allow much higher capacity of devices on integrated photonic circuits compared to the two-dimensional (2D) counterparts[23–25,31,32]. Interestingly, it has been experimentally demonstrated that a 2D microdisk cavity can be bent up to support WGMs exhibiting 3D trajectories in space, while keeping a considerably high quality factor[33,34]. It is therefore interesting to consider if a WGM defined on a curved 2D surface can be efficiently analyzed with both ray dynamics and wave optics, and most significantly, if such structures bring fundamental changes that would potentially leads to great improvement of device performance and on-chip integration.

In this article, we investigate the nature of photonic modes in asymmetric mcirocavities defined on a topologically curved surface. We develop a unique way of ray dynamics by modelling the light trajectories as the geodesic lines of the curved 2D space, and perform corresponding wave optics simulations to analyze the mode profile, optical loss and photonic interactions. We find that the cavity photon energy and lifetime can be continuously tailored by varying the 2D space curvature, leading to high quality factors (Q-factor) of the lossy chaotic and island modes, which are increased by up to ~200 times. Moreover, the space curvature brings modes of very different origins into resonance, resulting in the strong and weak coupling regimes, featured by the bifurcations of the photon energy and lifetime, respectively. By tuning the overall loss of the cavity, we reach an exceptional point of the non-Hermitian system at which the bifurcations of both the real and imaginary parts of the eigenvalue vanish. Our results combine the advances of optical chaotic dynamics, geodesic photonics and non-Hermitian physics, and would open the novel area of geodesic microcavity photonics for the improvement of device performance.

**The Model and Associated Definitions**

In a curved space, the traditional ray dynamics assuming straight light trajectories no longer applies, and the proper light trajectories are the geodesic lines determined by the space metric tensor, ensuring the optical length of extremum it passes. In comparison with flat 2D surfaces, curved 2D surfaces can be classified into two categories: the ones that can and cannot be unfolded to become a flat surface. The former, such as the bent-up microdisks[33], are not of our interest as its geodesic lines exactly correspond to the straight lines on the unfolded plane, making no difference in ray dynamics compared to a flat surface. Nevertheless, the latter is much more interesting, as it is formed based on a general definition of topological deformation from a flat surface by keeping the topological invariant. Such a surface reconstructs its own geodesic lines which generally lose the connection to the flat surface from which it is deformed, and thereby could potentially lead to completely new dynamics of light trajectories. It is sufficient to show such dramatic changes by simply building our 2D WGM cavity on a spherical surface, a type of simplest topologically curved 2D shape whose geodesic line is well known to be the great circle. Fig. 1**a** presented an illustrative picture of the cavity, in which the ROC represents the radius of curvature of the spherical surface, and the distance between any two points on the surface is defined by the length of the geodesic line between them. The ROB represents the radius of boundary of the cavity, i.e., the distance between the cavity boundary and the cavity geometrical center measured on the curved surface. The PSOS is derived by tracing the light trajectory as geodesic lines, which intersects with the cavity boundary, and reflects at the intersecting point back into the cavity following the law of reflection, i.e., equal incident and reflecting angle with the tangent of the boundary (see Methods for detail). Obviously, cavities with the same ratio $k =$ROB/ROC have exactly the same geometric shape. As the PSOS is irrelevant to the absolute size of the cavity, it is therefore the value of $k$ that determines how the space curvature affects the ray dynamics, and $k$ is then defined as the effective curvature. Fig. 1**b** and **c** show light trajectories of a symmetric circular cavity with $k=0.8$ (ROC=1, ROB=0.8) and its flat counterpart, respectively.

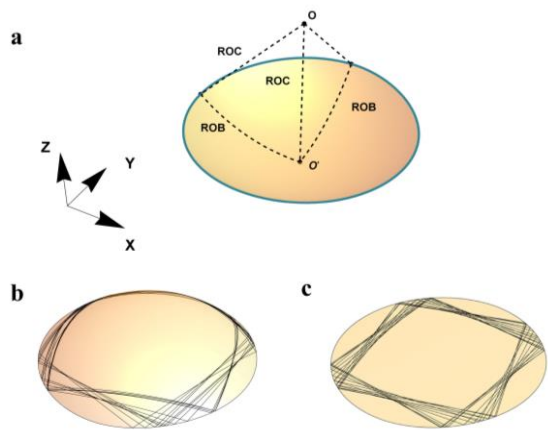

**Fig.1: WGM microcavities on topologically curved spherical surface. a** Schematic picture of the microcavity. O and O' are the center of the sphere and the geometric center of the cavity, respectively. ROC: radius of curvature of the spherical surface; ROB: radius of boundary of the cavity. **b** and **c** Light trajectories of a symmetric circular cavity with ROB=0.8 defined on a curved surface of ROC=1 (**b**) and a flat surface (**c**).

**Results and Discussion**

In the following we will show that the very simple change in space curvature can lead to giant difference in the optical mode properties. A symmetric circular cavity with a constant ROB on a spherical surface supports regular WGMs whose light trajectories form inscribed regular polygons, with all sides being the geodesic lines. The sinusoidal of the reflecting angles (noted as $\sin\chi$) of the WGMs associated with $N$-sided polygons is plotted as a function of the effective curvature $k$ in Fig. 2 **a**.

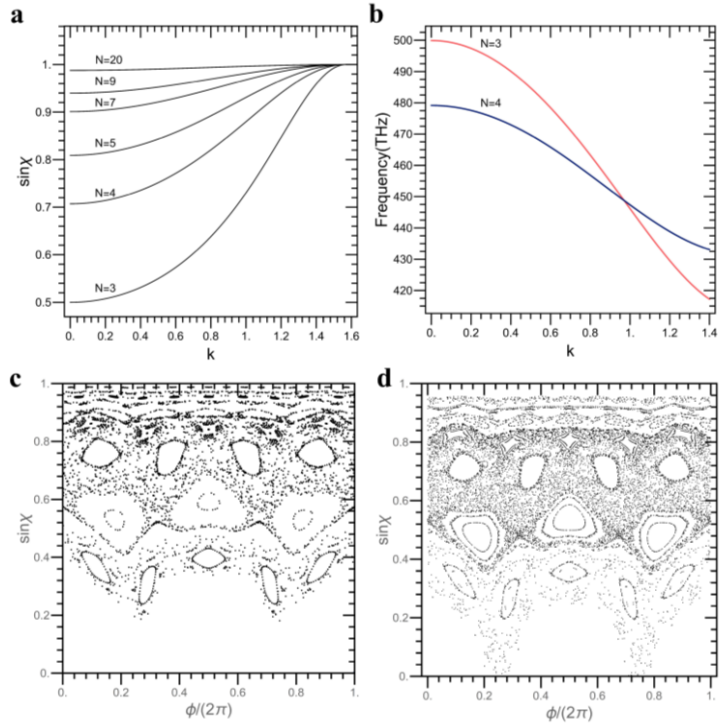

**Fig. 2: Ray dynamics of topologically curved microcavities. a** The sinusoidal of the reflecting angle $\sin\chi$ of the $N$-period modes as a function of the effective curvature $k$=ROB/ROC in a symmetric circular microcavity, showing an increase of $\sin\chi$ with k, with a larger amount of increase for smaller $N$. **b** Calculated resonant frequencies for the 3-period (red) and 4-period (blue) modes as a function of $k$ in a symmetric circular microcavity, showing different slopes and a crossing point. **c** and **d** The PSOS of a Face cavity defined on a curved surface with $k=\pi/6$ (**c**) and on a flat surface (**d**). The curved cavity shows non-uniformly up-shifted features with $\sin\chi$ compared to the flat one.

It is clearly seen that $\sin\chi$ increases with $k$, originating simply from the nature of the spherical surface on which the inner angle of a regular polygon is larger than that of its flat counterpart. Particularly, at $k=0$ (ROC is infinite) the cavity reduces to its flat counterparts, and at the extreme case of $k=\pi/2$ where the cavity becomes a surface of hemisphere with its boundary itself being a great circle, the WGMs reach a special regime where the trajectories of *N*=2 WGMs can exhibit arbitrary reflection angles and the trajectories of all the *N*>2 WGMs must be the cavity boundary itself, indicating a definite condition of sinχ = 1. Herein we constrain our studies in the regime of 0<*k*<π/2 to ensure no entire great circle is included, otherwise the cavity would work as a sphere cavity which support ordinary 2D WGMs.

The increase of sinχ with *k* indicates that the loss of the optical modes, especially those associated with small *N*, could be greatly reduced in curved space. Moreover, the space curvature pushes the light trajectories towards the interior of the surface, resulting in a variation of optical length of the round trip and thereby the resonant frequency. Fig.2**b** shows the calculated resonant frequencies for the 3-period (meaning light reflecting three times within one round trip, i.e. *N*=3, and thereafter) and 4-period modes as a function of *k*, derived by letting the optical length of the round trip equal to an integer time of half of the optical wavelength, namely,

$$v = m \times \frac{c}{2n_{eff}l} \qquad (1)$$

Where $v$ is the mode frequency, $n_{eff}$ is the effective refractive index, $l$ is the line length of the round trip given by the ray dynamics, $c$ is the vacuum light velocity and $m$ (integer) is the mode azimuthal order. Interestingly, the frequencies of the 3-period and 4-peirod modes change with *k* with different slopes, and become degenerate at *k*=0.9722. This kind of behavior can lead to the coherent coupling between modes of different physical origins, allowing the modification of the real and imaginary parts of the eigenvalues of the optical resonances.

To test the idea in detail, an asymmetric cavity needs to be designed to allow stable *N*-period island modes and the nearby chaotic modes, the Q-factor of which would be very low if the cavity is flat. We choose the well-known "Face cavity" [17] as an example to demonstrate the novel physics induced by the curved 2D space. The boundary of a "Face cavity" is expressed by

$$R = \begin{cases} R_0 \cdot (1 + a_2 \cdot \cos^2 \phi + a_3 \cdot \cos^3 \phi) & \phi \in \left(\frac{-\pi}{2}, \frac{\pi}{2}\right) \\ R_0 \cdot (1 + b_2 \cdot \cos^2 \phi + b_3 \cdot \cos^3 \phi) & \phi \in \left(\frac{\pi}{2}, \frac{3\pi}{2}\right) \end{cases} \qquad (2)$$

in which $\phi$ is the azimuthal angle with $\phi = 0$ pointing along the positive direction of the x-axis illustrated in Fig. 1. $R_0, a_2, a_3, b_2$ and $b_3$ are the shape parameters. The only difference from the flat cavity is that herein $R$ is the ROB measured along the geodesic line of the spherical surface. In our calculations we apply the shape parameters $R_0 = 1, a_2 = -0.1329, a_3 = 0.0948, b_2 = -0.0642, b_3 = -0.0224$. The PSOS of the "Face" cavity defined on a spherical surface of $k=\pi/6$ and its flat counterpart are shown in Fig. 2**c** and **d**, respectively. Whilst the details of some islands (e.g. the 5-peirod ones) and the distribution of chaotic modes become different, the most obvious feature is that all the island and the WGM modes are moved upwards by the space curvature to larger $\sin\chi$, compared to the flat cavity.

We perform mode analysis with wave optics using COMSOL Multiphysics (Version 5.6). A 3D space of simulation is established to accommodate the curved 2D cavities. We apply the "Eigenfrequency" module to obtain the frequencies, the Q-factors and the field distribution of all optical modes. The cavity parameters are set to be refractive index $n=\sqrt{10} \approx 3.1622$, cavity layer thickness $d=50$ nm. In the simulations, we keep the value $R_C = ROC * \sin(k)$ a constant of $2.1\mu m$, while varying the value of $k$ continuously from 0 to 1.4. One can verify from the 3D geometry that the circumference of the cavity is $2\pi R_C$, the constant value of which ensures the comparability among situations of different $k$. The value of ROB derived from $R_C$ and $k$ is then the value of $R_0$ in Eq. (2), following which the shape of the curved cavity is completely defined. We analyzed the 4-period island modes, the 3-period island modes and their nearby chaotic modes in the PSOS which are recognized by the simulation module, and plot their frequencies as a function of $k$ as colored dots in Fig. 3**a**, in which the colored solid lines represent a coarse fitting of modes using Eq. (1). The modes are labeled by the letters a-h, with the spatial profiles of light intensity and light trajectories of ray dynamics shown in Fig. 3**c**. The graphs are all presented in a top-down view, whilst the 3D view is available in the supplementary. It should be noted that even without the fitting lines, we can still easily identify and trace each mode of a-h thanks to the continuous variation of the mode frequencies and spatial features with $k$. By comparing the mode spatial profile, the light trajectories and the distribution on the PSOS (see supplementary information), we can clearly identify that mode a is a 4-period island mode, modes b and e are 3-period island modes, and modes d,g and h (resp. c and f) are chaotic modes appearing around (resp. below) the 3-peirod island modes in the PSOS.

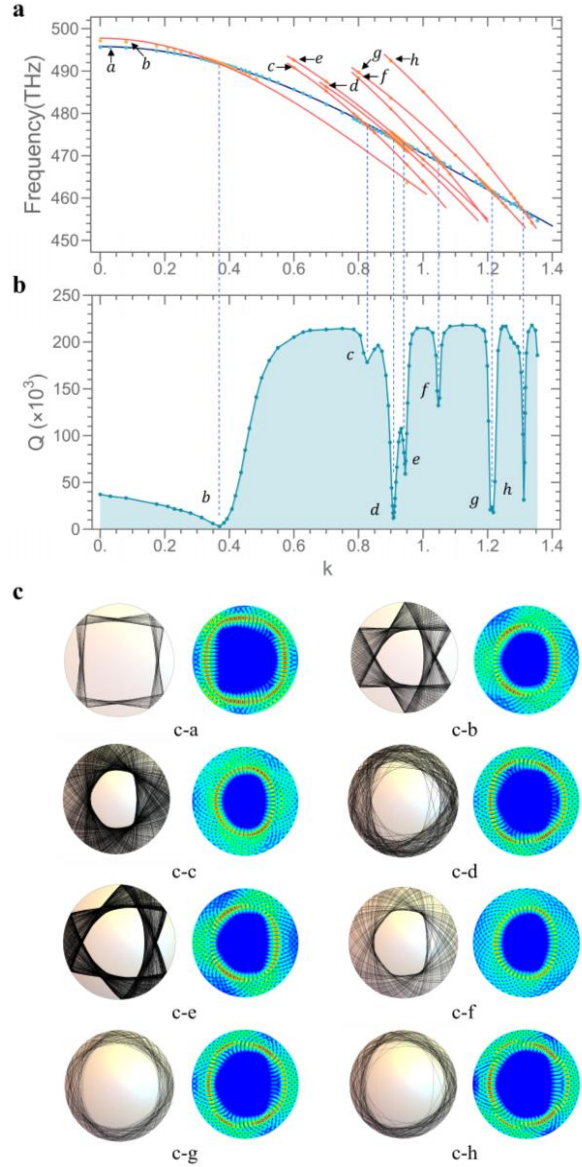

**Fig. 3: Features of various photonic modes in topologically curved microcavities. a-h labels modes of different origins. a** The frequency of modes a-h as a function of *k*, showing resonant points between different modes at different *k*. Dots are simulated values with wave optics and solid lines are coarse fitting using Equation (1). **b** The Q-factor of mode a (which is a 4-period mode) as a function of *k*, showing a general increasing trend and a series of dips at the resonant points with other low-Q modes (labeled with the corresponding letters). **c** Light trajectories [left panels of (c-a) to (c-h)] and electric field spatial profiles [right panels of (c-a) to (c-h)] of modes a-h (top view), with obvious similarities between the left and right panels, indicating the origins of each mode with the help of the corresponding PSOS (see supplementary information). Each graph in **c** corresponds to the point in **a** pointed by the arrow with the associated letter.

With increasing *k*, the modes of different origins show different slops of frequency variation, and crosses each other at certain points where they coherently interact. Indeed, after plotting the Q-factor of the 4-period island mode a as a function of *k* in Fig. 3**b**, we observe sharp dips at the crossing points with the 3-period island and chaotic modes which are more lossy, indicating coherent coupling between modes. On the other hand, if we ignore those dips, the Q-factor of the 4-period island mode generally increases with *k* up to a value of ~220,000, which is 4-5 times larger than the same mode of a flat Face cavity (~50,000) and comparable with the WGMs of a flat cavity. For the 3-period island mode b which exhibits even lower Q ~1,000 in a flat cavity, the Q factor still reaches ~220,000 at large *k* (see supplementary material), indicating an increase of ~200 times. These results robustly demonstrate the key merit of the space curvature that it reconstructs the light trajectories and reduces the cavity loss.

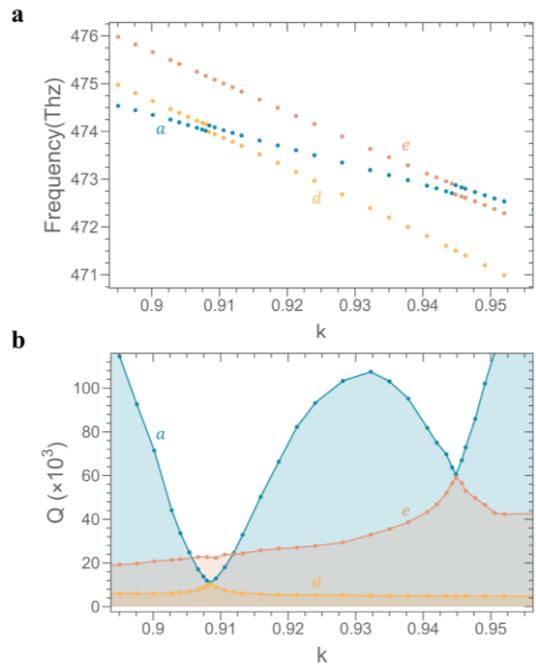

**Fig. 4: Details around the resonant points between modes a (blue) and d (yellow), and modes a and e (orange). a** The mode frequency as a function of k, showing anticrossing features at each resonant point. **b** The Q-factor of mode a as a function of k.

To further explore the mode coupling effect, we did simulations with smaller footsteps at the crossing point between modes a, d and e, as shown in Fig. 4, in which modes a, d and e are colored by blue, yellow and orange, respectively. Instead of simply crossing each other, the modes show anticrossing feature with a splitting of frequency at the resonant point, at $k=0.9081$ for a-d and $k=0.9444$ for a-e, where the Q-factors of the coupled two modes become identical. Frequency splitting and loss modification as a result of mode coupling has been investigated in WGM microcavities[4,20,35] and nanostructures[36], being a typical feature of the non-Hermitian system, which can be described by the Hamiltonian in matrix form:

$$\begin{pmatrix} \omega_1 - i\gamma_1 & \mu \\ \mu & \omega_2 - i\gamma_2 \end{pmatrix} \quad (3)$$

where $\omega_{1,2}$ and $\gamma_{1,2}$ are the mode frequency (the real part of the eigenvalue) and linewidth (the imaginary part of the eigenvalue) of the uncoupled bare modes and $\mu$ is the coupling strength. At the resonance condition of $\omega_1 = \omega_2 = \omega$, the eigenvalues of the coupled system are

$$\sigma_\pm = \omega - i\gamma_{ave} \pm \sqrt{\mu^2 - \gamma_{diff}^2} \quad (4)$$

where $\gamma_{ave} = \frac{\gamma_1 + \gamma_2}{2}$ and $\gamma_{diff} = \frac{\gamma_1 - \gamma_2}{2}$. When $\mu^2 - \gamma_{diff}^2 > 0$ (resp. $\mu^2 - \gamma_{diff}^2 < 0$) the system is in the strong (resp. weak) coupling regime featured by a bifurcation of the real (resp. imaginary) part of the eigenvalue. As the frequency and the Q-factor are associated with the real and imaginary parts of the eigenvalue respectively, the a-d and a-e mode couplings featured in Fig. 4 are both in the strong coupling regime. They can possibly turn into the weak coupling regime by modifying either the coupling strength $\mu$ and the difference in loss $\gamma_{diff}$. Indeed, changing the refractive index $n$ is an effective way to tune the difference in loss, as it has more influence on the lossy mode than the conservative mode. We perform mode analysis using COMSOL on a cavity of $k=0.9081$ where the bare modes a and d are in resonance, and vary $n$ with a small step size of 0.17. The results of the normalized frequency splitting (real part of $\frac{\sigma_+ - \sigma_-}{2}$) and the normalized linewidth difference (imaginary part of $\frac{\sigma_+ - \sigma_-}{2}$), shown as colored dots in Fig. 5a and b, exhibit bifurcation of the frequency (resp. the linewidth) when above (resp. blow) $n=2.88$, indicating the strong (resp. the weak) coupling regimes. At $n=2.88$, the non-Hermitian exceptional point (EP) connecting the strong and weak coupling regimes is reached, featured by the degeneracy of both the real and imaginary parts of the eigenvalues. These results can be well fitted by Eq. (4) with the parameter $\mu = 0.07$, shown as the solid lines. Points $(a_1,a_2)$ $(b_1, b_2)$ and $(c_1,c_2)$, which are in the weak coupling regime, at the EP and in the strong coupling regime are chosen to compare the spatial field distribution in Fig. 5c. Whilst exhibiting differences in both the weak and strong coupling regimes, the mode profiles are completely identical at the EP.

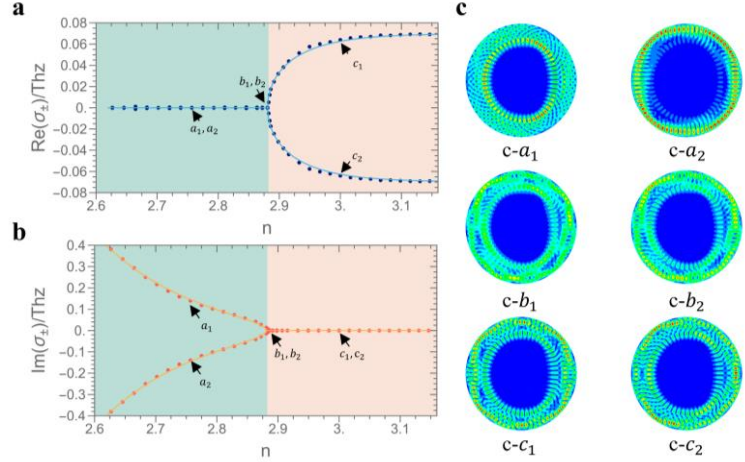

**Fig. 5: Non-Hermitian features and the exceptional points (EP). a** and **b** The real (**a**) and imaginary (**b**) parts of the eigenvalues of the hybrid modes formed by the coupling between modes a and d, as a function of the material refractive index *n*. **c** The electric field distributions (top view) of the corresponding points $a_1$, $a_2$, $b_1$, $b_2$ and $c_1$, $c_2$ in **a** and **b**. The degenerate points $b_1$ and $b_2$ corresponds to the EP. Note here the letters $a_1$ to $c_2$ are used to label the points in this figure and are not associated with the labeling of different modes.

It is interesting to look back at the Q-factor of the 4-peirod island mode in Fig. 3**b**. It seems the dips at the intersections of a-c and a-f are shallower than the others. This is because mode a is actually in weak coupling with the chaotic modes c and f, which exhibit the highest loss, even at a high refractive index of $n=\sqrt{10}$. As a result, the Q-factor bifurcates and the dips only show the higher values of Q which is associated with mode a. On the contrary, the other dips are in the strong coupling regime where the coupled two modes must share an identical Q, which is the averaged value and thereby much lower. The a-e dip also seems shallow despite the strong coupling between the two modes, owing to the lower loss associate with mode e which is a 3-period regular mode.

Finally, it is essential to give an interpretation to the physical meaning and possible applications of the curved microcavity and its non-Hermitian features. In addition to using the island and chaotic modes as dynamic channels to reach the high-Q WGMs [21,37], our results draw the attention to the practical value of these modes themselves. As shown in the field spatial profiles, the high-Q optical modes in a curved space, which can possibly be the low-ordered island ones, are not necessarily located only at the cavity periphery, but can go deep into the cavity area. This would allow the design of devices utilizing any given position of the cavity area that the photons pass, and would even enable vertical laser emission from the interior of the cavity area by modifying local geometries, e.g., inducing a shapely curved spot at the field maximum. Such manipulations are especially promising with the possibility of engineering the paths in the PSOS on demand[38]. When pumping to provide gain in the cavity, the position of EP would move according to the pumping intensity and spatial profile, which would potentially be used to design pumping-sensitive lasers and optical switches. The sharp variation of Q factor and frequency dispersion in the strong coupling regime implies good opportunities for sensing associated with shape and refractive index variations.

**Conclusion**

In conclusion, we investigated the dynamics of optical modes in topologically curved asymmetric microcavities, and reveals a series of novel effects including reduced loss by space curvature, geodesic mode coupling and non-Hermitian EP. We notice that the work is only based on the simplest shape of curvature, implying that much richer microcavity physics with space curvature are to be explored in the future. The basic principle of ray dynamics is generally applicable for any type of curved surface, while solving the geodesic equation would require advanced mathematical skills. Experimental realization of curved cavities could be potentially achieved via material strain, sophisticated lithography, laser milling and high resolution 3D printing, etc., whilst curved optical devices such as geodesic lenses have already been realized with soft matter[39,40]. The curved space-time in confined system could also be potentially interesting for optical simulators of astrophysics and cosmology[41–43]. With all the above important interests, our findings will initiate the novel area of geodesic microcavity photonics which would contribute in both new physics and practical advances of optoelectroinc devices.

**Methods**

**The algorithm of ray dynamics on curved surface.** We regard the curved 2D surfaces as very thin waveguides in which light travels with a uniform effective refractive index $n_{\text{eff}}$. Therefore, like travelling in straight lines in flat 2D cavities, light ray travels in geodesic lines in the curved space, which ensures the optical length of extremum it passes. The light trajectory of geodesic line intersects with the cavity boundary, and reflects at the intersecting point back into the cavity following the law of reflection, i.e., equal incident and reflecting angle with the tangent of the boundary. The algorithm thus contains 6 steps:

(1) Choose an arbitrary starting point $P_0$ at the cavity boundary (which is generally a spatial curve C) and an arbitrary initial traveling direction $l_0$ ( a tangent vector of the curved cavity surface Σ at $P_0$) ;

(2) Derive the equation of the geodesic line $S$ with $P_0$ and $l_0$. $S$ is then the trajectory of the incident beam.

(3) Derive the intersecting point $P$ between the incident beam $S$ and the cavity boundary $C$, by simultaneously solving the equations of both. $P$ is then the reflecting point.

(4) Derive the tangent vector of $S$ at $P$, noted as vector $l$.

(5) Derive the tangent vector of $C$ at $P$. noted as vector $t$. The angle formed by $l$ and $t$ is therefore the complementary angle of the incident angle at $P$.

(6) Derive the tangent vector of the reflected beam at $P$, noted as $l'$, following the law of reflection that the incident and reflecting angles are equivalent (so are their complementary angles), i.e., $l' \cdot t = l \cdot t$. Meanwhile, the reflected light has to be inside the cavity surface Σ, i.e., $l'$ has to be tangent to Σ, which requires $l' \cdot n = 0$, where $n$ is the normal vector of Σ at $P$. For spherical surfaces, $n$ is the unit vector along the radial directions of the sphere.

(7) Using $P$ and $l'$ as the new starting point and initial direction vector, run steps (2) –(6) repeatedly. After enough times of iterations, choose again a new set of $P_0$ and $l_0$ for a new round of calculation.

In the algorithm, all tangent and normal vectors can be calculated using standard mathematical tools of vector analysis. Although the derivation of geodesic line equations in general requires tensor analysis (thereby differentially geometry), for the specific situation of regularly shaped surfaces like the spherical ones, existing solutions of mathematical expressions are available, i.e., the great circle. In each iteration of the algorithm, the position $P$ and direction vector $l$ are recorded. The collection of $l$ vs. $P$ constitutes the PSOS which is the standard analyzing method also for flat 2D cavities.

**The COMSOL simulation.** The simulation is performed in a 3D cylindrical space with a radius of 3.6μm, a height of 5.1μm, a background refractive index of 1 (air) and absorbing boundaries. The simulation mesh is generated by the software, showing ~300 grids around the cavity periphery. In all simulations, the complex refractive index of the cavity material has a small imaginary part of $-7.9056 \times 10^{-6}$.

**Data availability**

The datasets generated and analyzed during the current study are available from the corresponding author on reasonable request.

**Acknowledgement**

This work is supported by National Natural Science Foundation of China (12074303, 11804267 and 11904279) and Shaanxi Key Science and Technology Innovation Team Project (2021TD-56).

**Conflict of interests**

The authors declare no conflicts of interest.

**Supplementary information** is available for this paper at url that will be provided by the publisher.

# Supplementary information for "Geodesic photon coupling and non-Hermitian exceptional point of microcavities on topologically curved space"

Yongsheng Wang, Xiaoxuan Luo, Bo Li, Zaoyu Chen, Zhenzhi Liu, Fu Liu, Yin Cai, Yanpeng Zhang, and Feng Li[*]

*Key Laboratory for Physical Electronics and Devices of the Ministry of Education & Shaanxi Key Lab of Information Photonic Technique, School of Electronic Science and Engineering, Faculty of Electronic and Information Engineering, Xi'an Jiaotong University, Xi'an 710049, China*

Supplementary Figures:

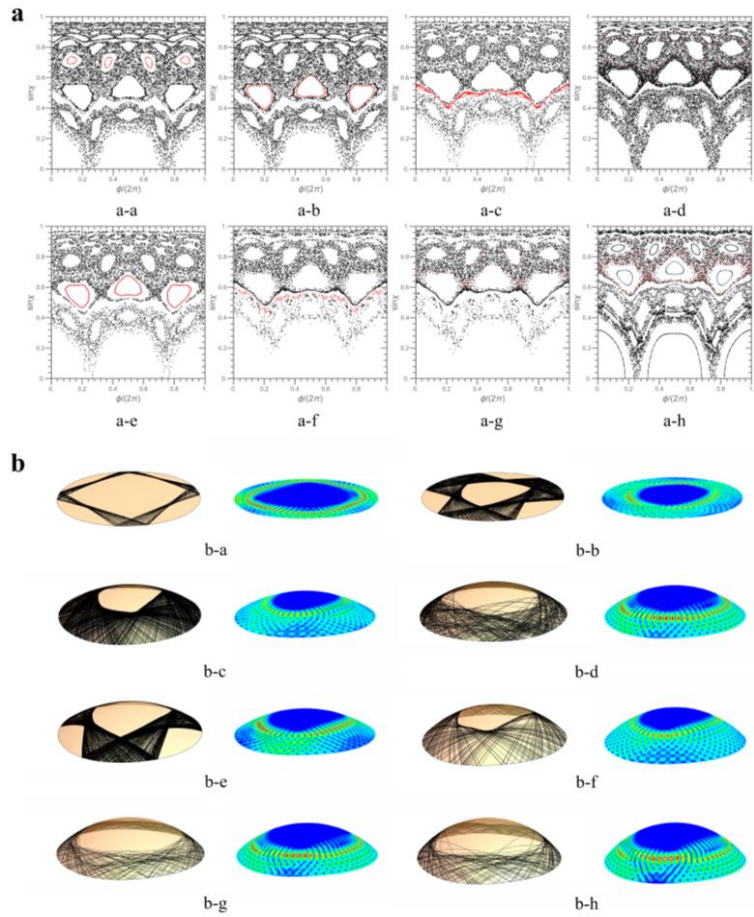

**Fig. S1: Ray Dynamics and wave optics simulations of modes a-h in Fig. 3 of the main text. a** The PSOSs of the curved microcavities supporting modes a-h [graphs from (a-a) to (a-h)], while the points corresponding to the trajectories in Fig. 3**c** of the main text are colored red, showing clearly the origin of each mode. **b** The side views of the light trajectories [left panels of (b-a) to (b-h)] and electric field distributions [right panels of (b-a) to (b-h)] of modes a-h.

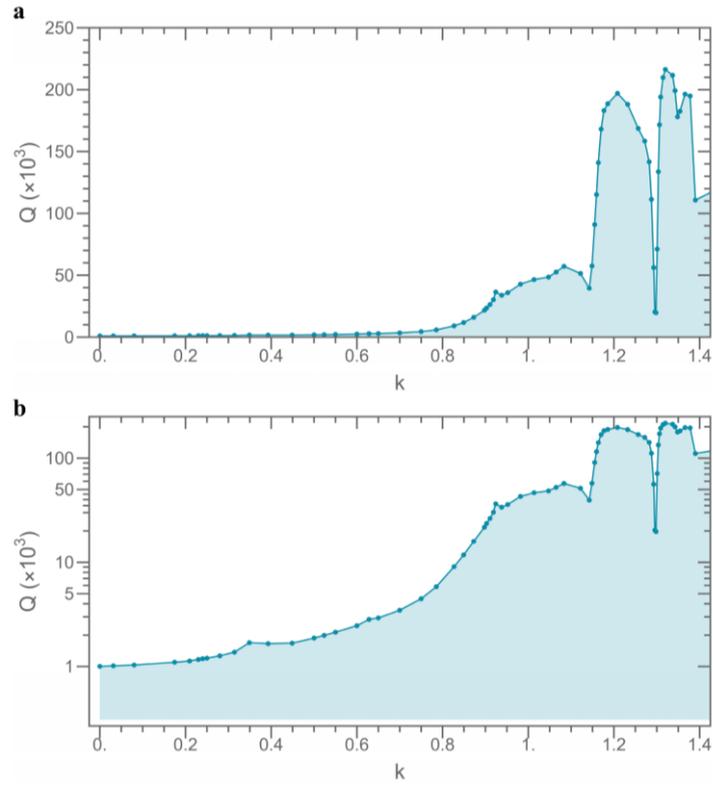

**Fig. S2:** The Q-factor of mode b, which is a 3-period island mode, as a function of $k$, showing that the Q-factor can be increased by more than 200 times by the space curvature. **a** and **b** are the same data plotted in linear and log scales, respectively.